\documentstyle[11pt,aaspp4]{article}

\begin{document}

\title{The Mass of Abell 1060 and AWM 7 from Spatially
Resolved X-ray Spectroscopy:
Variations in Baryon Fraction}
\author{M. Loewenstein\altaffilmark{1} and 
R. F. Mushotzky}
\affil{Laboratory for
High Energy Astrophysics, NASA/GSFC, Code 662, Greenbelt, MD 20771}
\altaffiltext{1}{Also with the Universities Space Research Association}

\begin{abstract}
Using X-ray temperature and surface brightness
profiles of the hot intracluster
medium (ICM) derived from {\it ASCA} and {\it ROSAT} observations we
place constraints on the dark matter (DM) and baryon fraction
distributions in the poor clusters Abell 1060 (A1060) and AWM 7.
Although their total mass distributions are similar, AWM 7
has twice the baryon fraction of A1060 in the best-fit models.
The functional form of the DM distribution
is ill-determined, however
mass models where
the baryon fractions in A1060 and AWM 7 significantly overlap
are excluded.
Such variations in baryon fraction are not predicted by standard
models and imply that some mechanism in addition to gravity plays a
major role in organizing matter on cluster scales.
\end{abstract}


\section{Introduction}

If gravity is the sole mechanism for organizing matter
on large scales, then clusters of galaxies should share a common ratio
of baryonic to nonbaryonic mass -- that of the universe as a whole.
Departures from such uniformity can be searched for
using X-ray
observations of the kind described in this paper; if such variations are
discovered
they provide direct evidence against the adequacy of
standard cosmogenic scenarios.
The poor clusters A1060 and AWM 7 comprise an excellent pair of
objects for a study of this type. They are nearby, bright clusters
of similar richness and X-ray temperature; however, AWM 7 is 
approximately five times more luminous in X-rays than A1060
(\cite{e91}).
This suggests that the fraction of the total mass
in baryons in AWM 7 is larger; this is confirmed
by the detailed analysis of recent high quality
X-ray data that follows.

\section{{\it ASCA} Temperature Profiles}

As required to 
constrain their DM distributions, 
accurate temperature profiles from
{\it ASCA}
spectra are available for A1060 and AWM 7. For 
A1060 we adopt the temperatures measured by \cite{t96},
while for AWM 7 the profile is obtained from archival
performance verification phase data. Spectra were extracted
in five annuli using standard screening and
background subtraction procedures (e.g., \cite{mus96}), and
fit with \cite{r77} thermal plasma models using the XSPEC package.
The observed profiles, with
90\% confidence uncertainties, are shown in Figures 1a and 1b.
The radial scale is computed for $H_o=50$ km s$^{-1}$
Mpc$^{-1}$, the value assumed throughout this paper.

The {\it ASCA} data have also been 
analyzed by \cite{m96},
who derive an identical temperature profile and note
the consistency with the {\it Einstein} MPC and {\it EXOSAT}.
{\it HEAO-1} A2 (\cite{s80}) and {\it Ginga} (\cite{t93})
large beam temperatures are consistent as well.
Temperatures derived from
{\it ROSAT} PSPC spectra are systematically lower 
by 20\% (\cite{ne95}, hereafter NB), 
perhaps due to small PSPC spectral calibration
problems (\cite{m96}). Our analysis is not
sensitive to the cooling flow inside $\sim 2'$
detected with the PSPC. 

Our temperature profiles do not account for
the broad, energy-dependent wings of the {\it ASCA} point spread function
(PSF), since any effect
for these relatively cool clusters is small (\cite{o95}). 
A comparison of our AWM 7 temperature
profile with that of
\cite{m96} and our own experience with PSF corrections
indicates that our uncertainties may be
underestimated by as much as a factor of 2. This is further supported
by our analysis of AWM 7 that includes the effects of
the {\it ASCA} PSF with software
developed by Dr. K. A. Arnaud.
For this reason we will be rather lenient in choosing
acceptable mass models.

\section{Models and Parameters}

We derive DM distributions
using the method of \cite{l94} that
produces physical, mathematically well-behaved temperature distributions
based on an integration of the equation of hydrostatic equilibrium (HSE)
inward from infinity. The consensus from numerical simulations of
cluster formation (e.g., \cite{n95}; \cite{s96};
\cite{e96}) is that departures from HSE
are generally small and mass estimation operating
under the hydrostatic assumption accurate.

The total gravitating mass is assumed to
consist of galaxy, gas, and nonbaryonic components. 
(The DM is assumed to be entirely nonbaryonic; any contribution
from MACHOs would increase the inferred baryon 
fractions.) The baryonic parameters
are based on optical and X-ray observations, while
the DM distribution
is parameterized by a scale-length and normalization.
The two DM parameters and the pressure at infinity are
varied until the model temperature distribution, appropriately averaged
and projected, matches the profile 
derived from {\it ASCA} spectra. 
A model is deemed acceptable if it predicts
temperatures within $3\times$ the 90\%
uncertainties of the best-fits to the data in each annulus.
The temperature in these models can take on arbitrary
values outside of the limited {\it ASCA} field-of-view.
We eliminate models where the temperature
climbs to extreme values just outside the observed region
by rejecting pressure-dominated models where the boundary
pressure at the virial radius, $r_{vir}$, 
exceeds the average inside $r_{vir}$ calculated
using the mass-weighted temperature in the observed region.

This procedure for constraining the DM
distribution effectively accounts for systematic uncertainties
in the mass model from projection effects and limited
spatial resolution and field-of-view. Because of the superb accuracy of
PSPC surface brightness profiles and {\it ASCA} temperatures these
effects dominate any statistical uncertainties.

The gas density distribution is modeled using the `$\beta$ model'
-- $\rho_{gas}=\rho_{gas,o}(1+r^2/a_{gas}^2)^{-1.5\beta}$, the
galaxy distribution as the sum of two `King profiles' --
$\rho_{gal}+\rho_{cd}=\rho_{gal,o}(1+r^2/a_{gal}^2)^{-1.5}+
\rho_{cd,o}(1+r^2/a_{cd}^2)^{-1.5}$ -- to account for both
central galaxy and smoothed cluster galaxy components.
Our standard model for the DM
distribution uses the function
proposed by \cite{n95} that is an excellent
characterization for a wide range of cluster formation scenarios 
(\cite{c96}, hereafter CL),

\begin{equation}
\rho_{dark}=\rho_{dark,o}(a_{dark}/r)(1+r/a_{dark})^{-\alpha}
\end{equation}

\noindent
with $\alpha=2$,
although we have also experimented with other functional forms.

\begin{deluxetable}{cccccccc}
\footnotesize
\tablecaption{Gas And Galaxy Mass Distribution Parameters} 
\tablewidth{475pt}
\tablehead{
\colhead{Cluster} & \colhead{$a_{gas}$} &
\colhead{$\rho_{gas,o}$} &
\colhead{$\beta$} &
\colhead{$a_{gal}$} &
\colhead{$\rho_{gal,o}$} &
\colhead{$a_{cd}$} &
\colhead{$\rho_{cd,o}$}}
\startdata
A1060 & 94 & 9.5 & 0.61 & 160 & 5.5 & 2.6 & $8.6\ 10^4$ \nl
AWM 7\tablenotemark{a} & 102 & 13. & 0.53 & 190 & 5.0 & 2.0 & $3.0\ 10^5$ \nl
\enddata

\end{deluxetable}
 
Our fixed, baryonic model parameters are displayed in Table 1. For
AWM 7 we use the $\beta$ model parameters from NB that we 
have confirmed using archival PSPC and {\it ASCA} data,
and include the central cooling flow excess. 
The galaxy parameters
are taken from \cite{f95} -- their integrated optical
luminosity agrees well with \cite{b84} and
\cite{d95}. 
For A1060 the $\beta$ model parameters
are based on our own best-fit to the 0.4-2.0 keV PSPC surface brightness
profile extracted from archival data. 
Galaxy parameters for A1060 are derived from
de Vaucouleurs model fits to the light profiles of the
overall cluster (\cite{f88}) and central
galaxy (\cite{v91}), converting from
de Vaucouleurs radius to King model core radius by dividing by 12,
and normalizing according to the luminosities 
in \cite{e91} and \cite{v91}.
We assume mass-to-light ratios, in
solar units, of 5 in the visual band and 7 in the blue. 

\section{Results and Discussion}

\subsection{Baryon Fractions in Acceptable Mass Models}

Figures 1a and 1b show 
projected, emission-averaged model temperature distributions superimposed
on the {\it ASCA} temperatures for A1060 and AWM 7. 
The solid lines are the distributions for the
best-fit mass models, the dotted lines the 
distributions with the most compact DM
density profiles 
that meet the acceptance criterion,
and the dashed lines those with the most diffuse DM
distributions.
The corresponding baryon fraction distributions as a function
of radius in units of $r_{vir}$ are shown in Figure 2.
We have defined $r_{vir}$, following CL,
as the radius within which the average overdensity is 178:
$2\pm 0.5$ Mpc for A1060, and $2\pm 0.4$ Mpc for AWM 7. 
From $0.1r_{vir}$ to $r_{vir}$ 
the baryon fraction distributions are flat, 
with a maximum increase of $\approx 2$
over this decade in both clusters; there may be a modest
decline for A1060.
Despite
spreading beyond the radius of the last
temperature measurement ($\sim 0.25 r_{vir}$), the observationally
allowed baryon fraction distributions are disjoint at all 
$r>0.05 r_{vir}$. Total masses and baryon fractions 
within 0.5, 1, and 2 Mpc are 
displayed in Table 2. For the best-fit models
the total mass distributions in the two clusters are
very similar, but AWM 7 has twice the baryon fraction.
Galaxies account for $\sim 40$, 30, and 20\% of the baryons within
0.5, 1, and 2 Mpc in A1060; the corresponding ratios in AWM 7
are $\sim 30$, 20, and 10\%: the ratio of stars to gas falls with
radius in both clusters. Since abundances and optical luminosities
are comparable, the ratio of intracluster (IC) metals to optical light is
about twice as high in AWM 7.

\begin{figure}
\plottwo{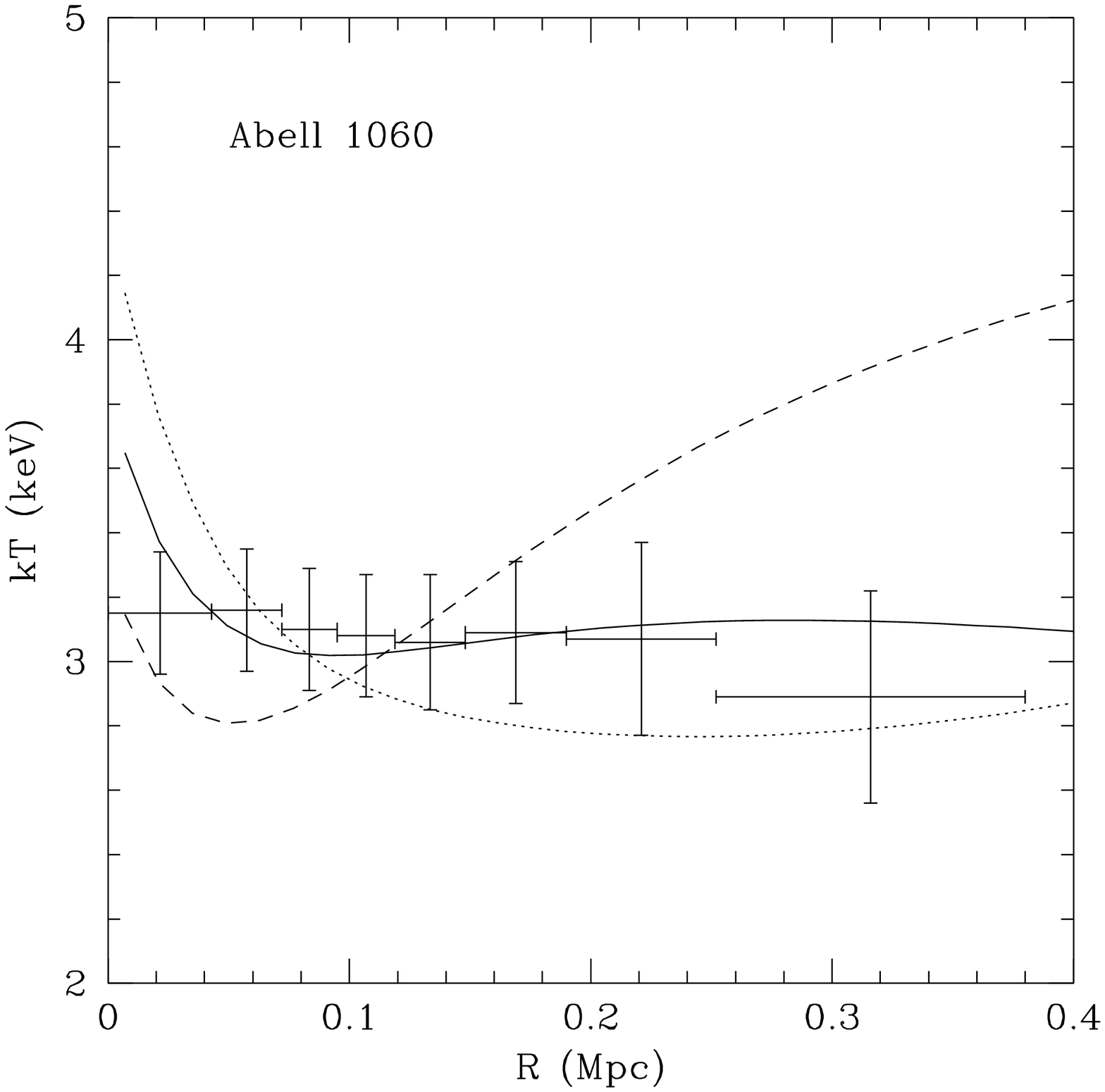}{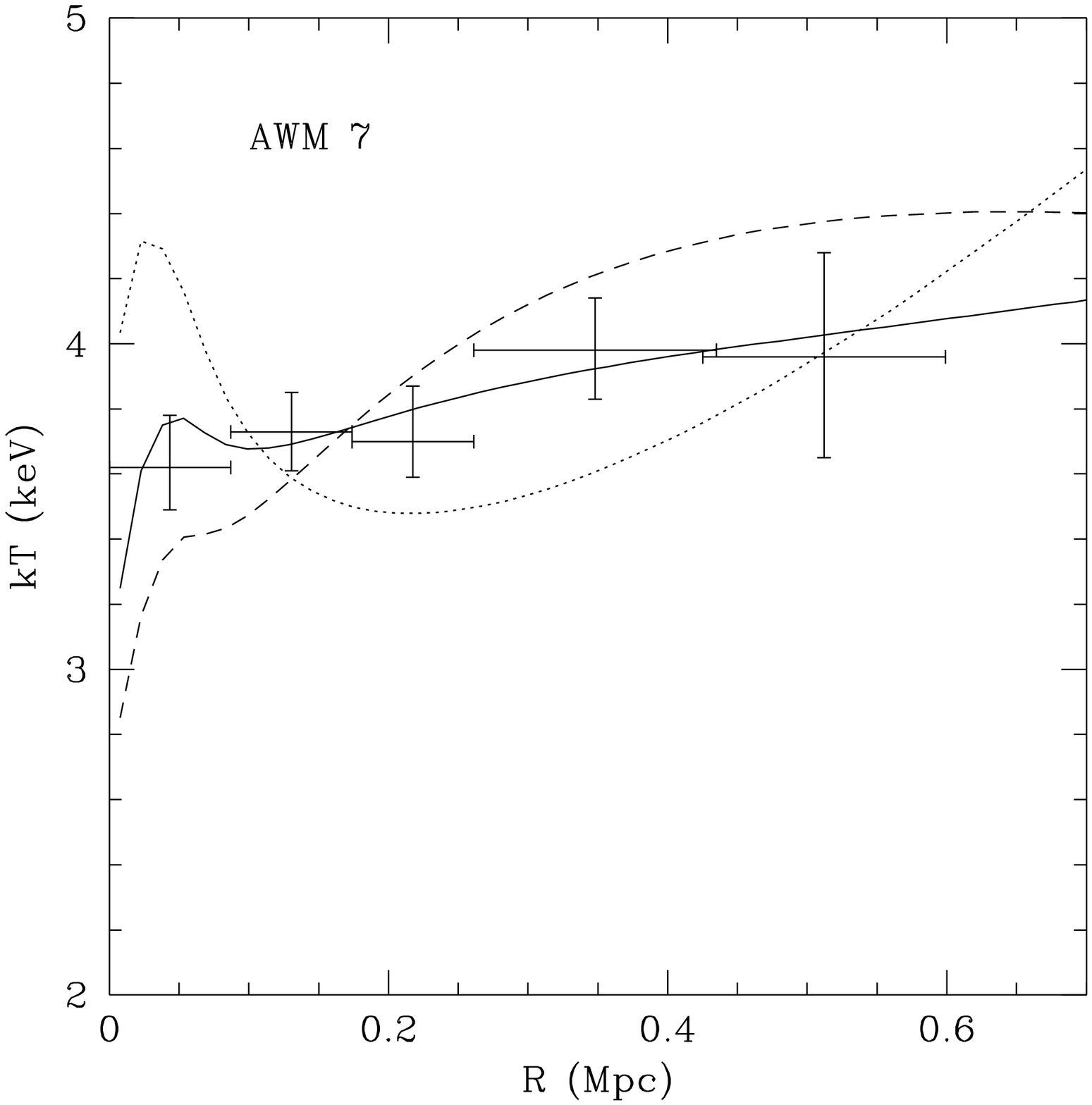}
\caption{(a) Projected, emission-averaged model temperature 
distribution superimposed
on the 90\% confidence limits from
{\it ASCA} (error bars) for A1060.
The solid line indicates the
best-fit mass model -- $a_{dark}=325$ kpc; the dotted (dashed)
line the most compact (diffuse) acceptable mass model --
$a_{dark}=150$ (900) kpc. (b) Same as (1) but for AWM 7 with
$a_{dark}=300$, 100, and 650 kpc for the best-fit, most compact, and most
diffuse dark matter models, respectively.}
\end{figure}

\begin{figure}
\plotone{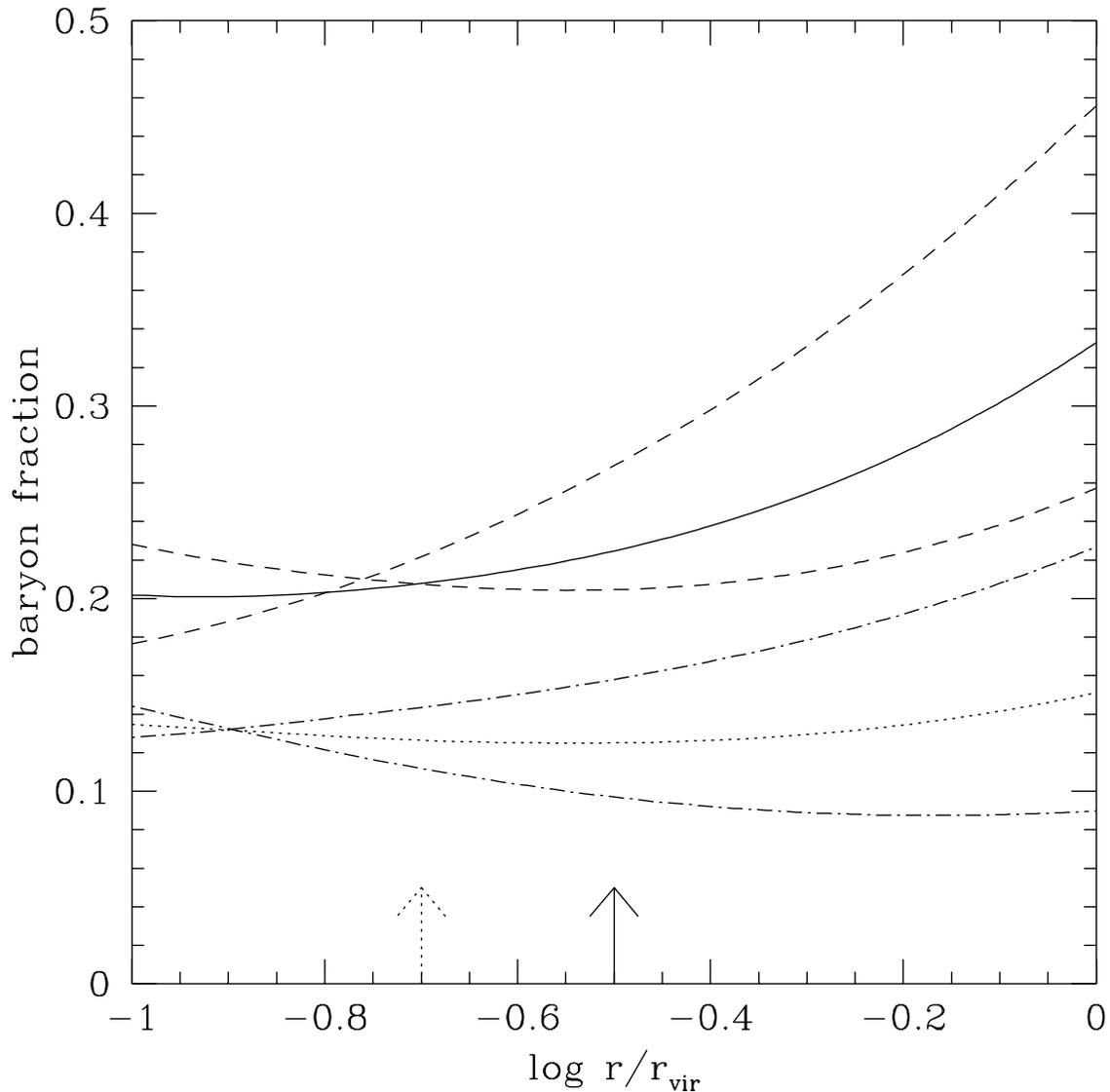}
\caption{Enclosed baryon fractions vs.
radius in units of the virial radius
in A1060 and AWM 7. The dotted (solid)
line represents the best-fit mass model, the
dot-dashed (dashed) lines the most compact and diffuse models
for A1060 (AWM 7). The compact models lie below the best fits.
Arrows at the bottom of the plot show the approximate
positions of the outermost
radii where {\it ASCA} temperatures have been obtained.}
\end{figure}

\begin{deluxetable}{cccccccc}
\footnotesize
\tablecaption{Total Masses And Baryon Fractions} 
\tablewidth{530pt}
\tablehead{
\colhead {Cluster} &
\colhead{${a_{dark}/r_{vir}}$\tablenotemark{a}} &
\colhead{$M(0.5)$\tablenotemark{b}} &
\colhead{$M(1.0)$} &
\colhead{$M(2.0)$} &
\colhead{$f(0.5)$\tablenotemark{c}} &
\colhead{$f(1.0)$} &
\colhead{$f(2.0)$}}
\startdata
A1060 & 0.17(0.09-0.36) & 1.0(0.80-1.1) & 2.1(1.4-2.9) & 3.6(2.2-6.3)
                         & 12(11-16) & 13(9-19) & 15(9-25) \nl
AWM 7 & 0.15(0.06-0.29) & 1.0(0.84-1.1) & 2.1(1.5-2.6) & 4.0(2.7-5.5)
                         & 21(20-26) & 25(21-36) & 33(25-50) \nl
\enddata

\tablecomments{Best-fit model total masses $M$ 
and baryon fractions $f$ evaluated within 0.5, 1.0, and 2.0 Mpc,
with allowed range in parentheses.}

\tablenotetext{a}{Ratio of dark matter scale length to virial radius.}
\tablenotetext{b}{Masses in units of $10^{14}M_{\odot}$.}
\tablenotetext{c}{Baryon fractions in percent.}

\end{deluxetable}

\subsection{Issues of Concern}

NB derive an acceptable range of baryon fraction in AWM 7 of
0.11-0.27 at 1.2 Mpc:
the lower limit
is significantly less than in any of our allowed models at
$\sim 1$ Mpc (Figure 2, Table 2). Examination of their Figure 6
reveals that the cause of the discrepancy is the sudden increase
in the upper envelope of their permitted mass profiles for $r>1$ Mpc.
Such models with arbitrary mass profile shapes are
allowed in NB, but
not in our work: we assume that the DM
can be characterized by
a single smooth function out to $r_{vir}$ (CL). 
The baryon fractions in NB are in good
agreement with ours inside 1 Mpc.

Although we include the departure from the $\beta$ model 
resulting from the cooling flow in AWM 7, we do not account for the
likely complex thermal structure. Therefore, our derived
baryon fractions inside $\sim r_{vir}/30$ may not be reliable in this
cluster.
Baryon fractions may be overestimated by as much as a factor of 2
in a multi-phase ICM (\cite{g96}). However,
neither multi-temperature fits to the {\it ASCA} spectra 
nor examination of the ratio of He-like to H-like Fe K lines
reveal any evidence for multiple temperatures in AWM 7 outside of
the cooling flow region, with a maximum
contribution (at 90\% confidence) from gas hotter than 6 kev of
20\%.

Finally, our assumption of circular symmetry cannot account for
the baryon fraction discrepancy between
A1060 and AWM 7 (\cite{w94}). 
If A1060 were prolate along the line-of-sight
its baryon fraction might be {\it overestimated} (\cite{d96});
such a geometry is unlikely in the case of AWM 7 which appears
elliptical in the plane of the sky.

\subsection{The Form of the Dark Matter Distribution}

X-ray observations of
the ICM in A1060 and AWM 7
are well-explained using a mass model where DM 
follows the universal function (Equation 1 with $\alpha=2$) 
predicted by numerical
simulations of
dissipationless hierarchical clustering in an $\Omega=1$ 
universe with gaussian initial
fluctuations. Furthermore, the constraints on the
scale-lengths ($a_{dark}/r_{vir}$; see Table 2) 
are consistent with cold dark matter
numerical experiments --
or other models with similar initial power spectra (\cite{n96}; CL).

However, because of the limited
spatial resolution and radial extent of the X-ray temperature profile,
alternative DM density distributions are also allowed.
DM models following Equation 1, but with $\alpha=1$ or 3, can
be constructed that provide fits to the data
of comparable quality to the standard $\alpha=2$ model.
The acceptable $\alpha=1$ ($\alpha=3$) models have correspondingly
smaller (larger) scale-lengths than their $\alpha=2$
counterparts. Models with cores also prove acceptable: we
cannot constrain the DM slope at small or large radii.

The inferred total mass within the outermost
radii where the temperature has been measured using {\it ASCA}
(380 kpc for A1060 and 610 kpc for AWM 7)
is independent of the assumed
DM density slope.
However, if the DM
distribution is
steep in A1060 the extrapolated baryon fraction could
increase significantly beyond 1 Mpc: in this case the 
baryon fractions in the acceptable models for A1060
and AWM 7 overlap for
$r>0.75r_{vir}$. If the DM slope were
steep in A1060 but flat in AWM 7 
this region of overlap could be extended inwards to $r>0.6r_{vir}$.
However, the baryon fraction distributions remain disjoint inside
1 Mpc and the discrepancy between the best-fit models is still a factor of 2
at all $r<r_{vir}$. The baryon fraction distribution
for all acceptable mass models with $1\le\alpha\le 3$ is fairly
flat, with maximum increases from $0.1r_{vir}$ to $r_{vir}$
of 2 and 3 
for A1060 and AWM 7, respectively, and a maximum decline
of a factor of 2 in A1060.

Finally we note that, regardless of the form of the DM
distribution,
the average projected total mass surface density is $\sim 0.15$ and
$\sim 0.09$ gm cm$^{-2}$ within 100 and 200 kpc for both 
clusters. This is well below the critical density for strong gravitational
lensing; however, these nearby clusters do not belong
to the class of extremely hot and luminous systems
where giant arcs have been detected at moderate redshift.

\subsection{Implications}

We have been conservative in constraining the
mass distributions in A1060 and AWM 7, since
the models with the smallest
DM scale-lengths require a substantial, fine-tuned
boundary pressure to reproduce the nearly flat observed
temperature profiles. Moreover, the temperatures in these models
rise linearly with radius beyond $\sim 500$ kpc
({\it i.e.} just outside the {\it ASCA}
field of view):
behavior not observed in more distant clusters. Thus the 
upper limit curves in Figure 2 are likely to be overestimates
(and may already be ruled out by temperature measurements 
at larger radii in AWM 7; NB, \cite{ez96}), and the allowed
range of baryon fractions
in A1060 and AWM 7 even more disjoint.

The X-ray data for A1060 and AWM 7 are best characterized
by models where the baryon fraction in AWM 7 is approximately twice 
that of A1060 at all radii from $0.05 r_{vir}$ to $r_{vir}$ --
about what one would expect from a simple scaling based on their
X-ray luminosities.
Under conservative assumptions about the uncertainties
and allowing for differences in the shape of
their DM
distributions, the
baryon fractions in A1060 and AWM 7 are marginally consistent at
$r\sim r_{vir}$, but are clearly
distinct inside $\sim r_{vir}/3$.
Although a wide range of baryon fractions can be found in the literature
(e.g., \cite{w95}), we believe this is the
first
cluster-to-cluster comparison to explore the uncertainties resulting
from different possible DM configurations
and to demonstrate the necessity of a range of cluster baryon fractions. To
the extent that X-ray temperature is a measure of total mass
(\cite{e96}) the spread in 
X-ray luminosity and mass accretion rate 
(two indicators of baryon richness)
at a given X-ray temperature also implies a range
in baryon fraction (\cite{f94}), 
although clusters must be individually
analyzed to verify this.

In standard
cluster formation models driven solely by gravitational instability
there is no mechanism for separating baryons and DM.
Variations in baryon fraction of the kind we are reporting thus
require an additional process. The most likely candidate
is feedback, in the form of powerful
supernovae-driven galactic outflows, from star formation in 
young galaxies during the protocluster epoch. Analysis of
{\it ASCA} spectra of a number of clusters indicates that, not
only is such powerful energy injection plausible, but is required
to account for 
measured ICM
abundances of type II supernova products such as O and Si
(\cite{mus96}, \cite{l96}).
In some clusters this energy input could have resulted in the ejection
of substantial amounts of IC gas into intercluster space, thus
reducing their baryon fractions. Variations in gas retention could result
from variations in star formation efficiency, initial mass
function, or differences in 
timing in the sense that in a less relaxed cluster mass loss is
facilitated by the occurrence of star formation 
in previrialized subclusters
with relatively small binding masses.

This mechanism should be less efficient for more massive clusters; and indeed,
preliminary analysis using {\it ASCA} temperatures of hotter clusters
shows a remarkable uniformity in baryon fraction (\cite{m95}).
This is also consistent with
the large apparent baryon fraction variations in groups (\cite{mu96})
where, moreover, the spread in the ratio of gas to stars may indicate that this
energy injection can have a negative
feedback effect on subsequent galaxy formation.

\section{Summary and Conclusions: Beyond the Baryon Catastrophe}

We have derived constraints on the DM and
baryon fraction distributions in the poor clusters A1060 and AWM 7
using {\it ASCA} and {\it ROSAT} observations of their
ICM. The data are consistent
with mass models where the form and length-scale of the 
DM distribution is identical to that found in numerical
simulations of cluster formation. However the spatial
resolution and extent of the observed temperature profiles is
not sufficient to rule out other DM models or 
determine whether the baryon fraction distribution is flat
or increasing (or decreasing, in the case of A1060) with radius.

Although they have similar optical luminosities and total masses
the total baryon fraction distributions in A1060 and AWM 7
are disjoint, with the X-ray data favoring models where AWM 7 is
approximately
twice as baryon rich as A1060 for $0.05r_{vir}<r<r_{vir}$.
The data for these and other clusters is consistent with a universal
baryon fraction of $\sim 0.25$ ($\Omega=0.1-0.4$, given nucleosynthetic
constraints on the overall baryon density), but requires the existence of
some gasdynamical
process to deplete the number of baryons in a fraction of relatively
low-mass systems (poor clusters and groups). Feedback from
galaxy formation is a likely candidate to provide such a mechanism and
is indicated by cluster abundance studies. Variations
in baryon fraction bring standard models
of structure formation where $\Omega=1$, standard Big Bang nucleosynthesis
is assumed, and gravity is the sole mechanism for organizing matter, into
further conflict with the observed universe.

\acknowledgments

We thank T. Tamura for providing us with the A1060
temperature profile, and K. Arnaud for making
spectral analysis software that includes the effects of the
{\it ASCA} PSF available to us and for feedback on the draft
manuscript.


\clearpage

\begin{thebibliography}{}
\bibitem[Beers et al. 1984]{b84}
Beers, T. C., Geller, M. J., Huchra, J. P., Latham, D. W., \& Davis, R. J. 
1984, \apj, 283, 33
\bibitem[Cole \& Lacey 1996]{c96}
Cole, S. M., \& Lacey, C. G. 1996, \mnras, 281, 716
\bibitem[Daines et al. 1996]{d96}
Daines. S., Jones, C., Forman, W., \& Tyson, A. 1996, preprint
\bibitem[Dell'Antonio et al. 1995]{d95}
Dell'Antonio, I. P., Geller, M. J., \& Fabricant, D. G. 1995, \aj, 110, 502
\bibitem[Edge \& Stewart 1991]{e91}
Edge, A. C., \& Stewart, G. C. 1991, \mnras, 252, 428
\bibitem[Evrard et al. 1996]{e96}
Evrard, A. E., Metzler, C. A., \& Navarro, J. F. 1996, \apj, in press
\bibitem[Ezawa et al. 1996]{ez96}
Ezawa, H., Fukazwa, Y., Haiguang, X., Ikuchi, K., Makishima, K., Ohashi, T.,
Tamura, 
T., \& Yamasaki, N. 1996, in X-ray Imaging and Spectroscopy of Cosmic Plasmas,
ed. F. Makino (Tokyo: Univ. Academy Press), in press
\bibitem[Fabian et al. 1994]{f94}
Fabian, A. C., Crawford, C. S., Edge, A. C., \& Mushotzky, R. F. 1994, \mnras, 
267, 779
\bibitem[Fitchett \& Merritt 1988]{f88}
Fitchett, M., \& Merritt, D. 1988, \apj, 335, 18
\bibitem[Fujita \& Kodama 1995]{f95}
Fujita, Y., \& Kodama, H. 1995, \apj, 452, 177
\bibitem[Gunn \& Thomas 1996]{g96}
Gunn, K. F., \& Thomas, P. A. 1996, \mnras, in press
\bibitem[Loewenstein 1994]{l94}
Loewenstein, M. 1994, \apj, 431, 91
\bibitem[Loewenstein \& Mushotzky 1996]{l96}
Loewenstein, M., \& Mushotzky, R. F. 1996, \apj, in press
\bibitem[Markevitch \& Vikhlinin 1996]{m96}
Markevitch, M., \& Vikhlinin, A. 1996, \apj, submitted
\bibitem[Mulchaey et al. 1996]{mu96}
Mulchaey, J. S., Davis, D. S., Mushotzky, R. F., \& Burstein, D. 1996,
\apj, 456, 80
\bibitem[Mushotzky et al. 1995]{m95}
Mushotzky, R. F., Loewenstein, M., Arnaud, K. A., \& Fukazawa, Y. 1995, in 
Dark Matter, ed. S. S. Holt \& C. L. Bennett (New York: AIP), p. 231
\bibitem[Mushotzky et al. 1996]{mus96}
Mushotzky, R. F., Loewenstein, M., Arnaud, K. A., Tamura, T., Fukazawa,
Y., Matsushita, K., Kikuchi, K., \& Hatsukade, I. 1996, \apj, in press
\bibitem[Navarro et al. 1995]{n95}
Navarro, J. F., Frenk, C. S., \& White, S. D. M. 1995, \mnras, 275, 720
\bibitem[Navarro et al. 1995]{n96}
Navarro, J. F., Frenk, C. S., \& White, S. D. M. 1996, \apj, 462, 563
\bibitem[Neumann \& B\"ohringer 1995]{ne95}
Neumann, D. M., \& B\"ohringer, H. 1995, \aap. 301, 865 (NB)
\bibitem[Ohashi 1995]{o95}
Ohashi, T. 1995, in Dark Matter, ed. S. S. Holt \& C. L. Bennett (New York: 
AIP), p. 255
\bibitem[Raymond \& Smith 1977]{r77}
Raymond, J. C., \& Smith, B. W. 1977, \apjs, 35, 419
\bibitem[Schindler 1996]{s96}
Schindler, S. 1996, \aap, 305, 858
\bibitem[Schwartz et al. 1980]{s80}
Schwartz, D. A., Davis, M., Doxsey, R. E., Griffiths, R. E., Huchra, J.,
Johnston, 
M. D., Mushotzky, R. F., Swank, J., \& Tonry, J. 1980, \apj, 238,
L53
\bibitem[Takahashi 1995]{t95}
Takahashi, T., Markevitch, M., Fukazawa, Y., Ikebe, Y., Ishisaki, Y., Kikuchi,
K., 
Makishima, K., \& Tawara, Y 1995, {\it ASCA} Newsletter, no. 3 (NASA/GSFC)
\bibitem[Tamura 1996]{t96}
Tamura, T., Day, C. S. R., Fukazawa, Y., Hatsukade, I., Ikebe, Y., 
Makishima, K., 
Mushotzky, R. F., Ohashi, T., Takenake, K., \& Yamashita, K. 1996, PASJ, 
in press
\bibitem[Tsuru 1993]{t93}
Tsuru, T. 1993, Ph D. Thesis, University of Tokyo
\bibitem[Vasterberg 1991]{v91}
Vasterberg, A. R., Jorsater, S., \& Lindblad, P. O. 1991, A\&A, 247, 335
\bibitem[White \& Fabian]{w95}
White, D. A., \& Fabian, A. C. 1995, \mnras, 273, 72
\bibitem[White et al. 1994]{w94}
White, D. A., Fabian, A. C., Allen, S. W., Edge, A. C., Crawford, C. S.,
Johnstone, R. M., Stewart, G. C., \& Voges, W. 1994, \mnras, 269, 589
\end{thebibliography}
\end{document}